\documentclass{edm_article}

\usepackage[protrusion]{microtype} 
\usepackage{enumitem} 
\setitemize{itemsep=0pt,topsep=0pt}
\setenumerate{itemsep=0pt,topsep=0pt}

\usepackage[breaklinks,colorlinks,linkcolor=blue,urlcolor=blue,citecolor=blue]{hyperref} 
\usepackage[nameinlink]{cleveref}
\usepackage{bm} 

\usepackage[normalem]{ulem}
\makeatletter
\g@addto@macro{\email}{\normalsize}
\makeatother

\begin{document}

\title{E2Vec: Feature Embedding with Temporal Information for~Analyzing Student Actions in E-Book Systems}

\numberofauthors{6}
\author{
\alignauthor
Yuma Miyazaki \\
    \affaddr{Kyushu University}\\
    \email{miyazaki.yuma.086@s.kyushu-u.ac.jp}
\alignauthor
Valdemar Švábenský \\
    \affaddr{Kyushu University}\\
    \email{valdemar.research@gmail.com}
\alignauthor
Yuta Taniguchi\\
    \affaddr{Kyushu University}\\
    \email{yuta.taniguchi.y.t@gmail.com}
\and
\alignauthor
Fumiya Okubo\\
    \affaddr{Kyushu University}\\
    \email{fokubo@ait.kyushu-u.ac.jp}
\alignauthor
Tsubasa Minematsu\\
    \affaddr{Kyushu University}\\
    \email{minematsu@limu.ait.kyushu-u.ac.jp}
\alignauthor
Atsushi Shimada\\
    \affaddr{Kyushu University}\\
    \email{atsushi@ait.kyushu-u.ac.jp}
}

\toappear{\scriptsize Y. Miyazaki, V. Švábenský, Y. Taniguchi, F. Okubo, T. Minematsu, and A. Shimada. E2Vec: Feature Embedding with Temporal Information for~Analyzing Student Actions in E-Book Systems. In \textit{Proceedings of the 17th International Conference on Educational Data Mining}, pages 434--442, Atlanta, GA, USA, July 2024. Editors: B. Paaßen and C. D. Epp. International Educational Data Mining Society.\\

© 2024 Copyright is held by the author(s). This work is distributed under the Creative Commons Attribution NonCommercial NoDerivatives 4.0 International (CC BY-NC-ND 4.0) license.
\\\url{https://doi.org/10.5281/zenodo.12729854}
}

\maketitle

\begin{abstract}
Digital textbook (e-book) systems record student interactions with textbooks as a sequence of events called EventStream data. In the past, researchers extracted meaningful features from EventStream, and utilized them as inputs for downstream tasks such as grade prediction and modeling of student behavior. Previous research evaluated models that mainly used statistical-based features derived from EventStream logs, such as the number of operation types or access frequencies. While these features are useful for providing certain insights, they lack temporal information that captures fine-grained differences in learning behaviors among different students. This study proposes E2Vec, a novel feature representation method based on word embeddings. The proposed method regards operation logs and their time intervals for each student as a string sequence of characters and generates a student vector of learning activity features that incorporates time information. We applied fastText to generate an embedding vector for each of 305 students in a dataset from two years of computer science courses. Then, we investigated the effectiveness of E2Vec in an at-risk detection task, demonstrating potential for generalizability and performance.
\end{abstract}


\section{Introduction}

Digital textbook systems are widely used in educational institutions and online educational services. They not only provide learning materials to students but also collect logs of student actions, such as moving between pages and adding makers or memos, in the EventStream format. Various studies have been conducted in the fields of Educational Data Mining (EDM) and Learning Analytics (LA), ranging from basic learning activity analysis to a deeper evaluation towards personalized learning support. In such analyses of e-book data, extracting useful features from EventStream is a crucial aspect since the features are used as inputs of downstream machine learning (ML) tasks, such as prediction of grades or clustering of students based on their behavior.

Previous studies (see \Cref{Feature Representation of Learning Activities}) used e-book EventStream data to extract features consisting of the counts of various actions (operations)~\cite{Okubo2017, chen2021predicting, akccapinar2019developing, Xing2018dropout}.
However, these types of features do not consider the sequential or temporal information of operations or the time intervals between operations. Flanagan et al. \cite{Flanagan2022Early} used features that consider sequential operations, taking into account the order of operations, but not the intervals between them. Therefore, it is not possible to capture in detail the differences between how students spent their time reading the learning materials.

Minematsu et al. \cite{minematsu2023const} proposed CRE (contrastive learning for reading behavior embedding), a method of feature embedding of EventStream data. They used operation tokens and timestamps as inputs of the embedding model. Subsequently, student vectors generated with CRE yielded higher F1-score than count-based features in downstream tasks. This model is the basis for our proposed method; however, we introduce a different unsupervised training model of embedding.

Our research purpose is to evaluate a more fine-grained representation of learning activities that considers not only the operations' order, but also their intervals during student learning. We propose a novel feature representation method for EventStream based on word embedding. Our method considers a set of operations and intervals between operations as $primitive$ symbols represented by individual characters. A short learning activity, such as opening an e-book and reading several pages by flipping them at certain time intervals, is represented as a series of $primitive$s, named $unit$. A sequence of $unit$s represents a learning activity, $action$, over time.

A feature embedding model based on fastText~\cite{bojanowski2016enriching}, proposed for effective word embedding, was trained using a large dataset of $action$s. Embedding features are then acquired using the trained model. We call the embedding model ``E2Vec'' since EventStream data are vectorized. The resulting embedding features are easy to use in various ML tasks.

E2Vec aims to improve downstream tasks in education by providing features that were not considered much in prior work. Our study empirically investigates whether features obtained by E2Vec are effective for downstream tasks, evaluating E2Vec on a task of predicting at-risk students. Specifically, we explore three research questions:
\begin{itemize}
    \item[RQ1] Does E2Vec represent learning activities in an appropriate way, meaning that (a) $unit$s similar to each other and (b) $action$s similar to each other are converted to a similar vector embedding?
    \item[RQ2] How well do features generated by E2Vec perform when used as inputs for downstream at-risk prediction tasks?
    \item[RQ3] How well do features produced by E2Vec generalize for applications in different models?
\end{itemize} 

\section{Related Work}
\subsection{Features Representing Learning Actions}
\label{Feature Representation of Learning Activities}

Past research employed EventStream data to extract features describing students' learning activities in various ways.

Okubo et al. \cite{Okubo2017} proposed the Active Learner Point (ALP) as a feature representation of student activities in e-books and other learning management systems. In ALP, learning activities, such as the number of markers and memos or attendance information, were scored in the range between 0 and 5. The scores were used for training a recurrent neural network (RNN) model for a grade prediction task.

Chen et al. \cite{chen2021predicting} researched early prediction of student grades using several ML classifiers. They focused on 14 kinds of operations in an e-book system, and extracted the number of each kind of operation. The result showed that operations with e-book memos positively correlated with the students' academic achievement.    

Ak{\c{c}}ap{\i}nar et al. \cite{akccapinar2019developing} used features such as the total number of events, total time spent in the e-book system, and the number of ``next page'' events. Additionally, they used the number of events longer than 3 seconds and shorter or equal to 3 seconds. These were the only considered features that involved time interval information between operations.

Flanagan et al.~\cite{Flanagan2022Early} used sequential features. They created 5-gram features from preprocessed logs and predicted the performance in an open-book assessment. However, time information was only used to determine whether an operation is recorded before or after starting the assessment, and operations with intervals of less than 3 seconds were removed in the preprocessing step. 

Another possibility of using EventStream data is a pattern analysis or clustering based on learning activities. Yin et al. \cite{yin2019exploring} performed k-means clustering of students based on learning activities and analyzed the characteristics of each cluster. Examples of features include total number of pages read and the frequency of page flipping. 

Overall, EventStream data can yield various features. A common issue among previous studies is that they did not fully exploit the potential of employing temporal information. This study aims to address this gap in the literature.

\subsection{Word Embedding}
\label{word embedding}

In Natural Language Processing (NLP), word embedding models are used to provide feature representations for words or documents.
Word2vec \cite{mikolov2013efficient} is one such well-known model, obtaining distributed representations of words in training text data.
GloVe \cite{pennington2014glove} and fastText \cite{bojanowski2016enriching} were proposed for an improved representation. 
FastText can generate a vector of words that does not exist in the training data using subword information. (A subword is an n-gram of characters.)
Recently, these models have been combined with other ML models for prediction and classification tasks -- see below.

Umer et al. \cite{umer2023impact} proposed a 3-layer Convolutional Neural Network (CNN) with fastText embedding for text classification and marked higher accuracy than other models on five datasets.  
Dharma et al. \cite{dharma2022accuracy} compared the three word embedding models -- Word2vec, Glove, and fastText -- using text classification with CNN. 
Furthermore, Riza et al. \cite{riza2021emotion} used a long short term memory network and fastText for emotion detection. 
Tiun et al \cite{Tiun_2020} showed the effectiveness of fastText in classification of functional and non-functional requirements in software engineering.  

Word embedding models were applied not only to NLP tasks but also in other fields. An example is item2vec by Barkan et al. \cite{barkan2016item2vec}. This model learned distributed representations of items that correspond to words in Word2vec. Moreover,~in bioinformatics, Ng et al. \cite{ng2017dna2vec} proposed dna2vec which is a method for feature generation using skip-gram.

To summarize, the value of feature embedding inspired by word embedding has been recognized in various studies. We propose a new feature embedding method based on the fastText model for EDM. Then, we verify its effectiveness for student learning logs from e-books. 

\section{The Proposed Method}
\label{Proposed method}

\begin{figure}[!htb]
\Description{Overview of the proposed method.}
\centering
\includegraphics[width=\linewidth]{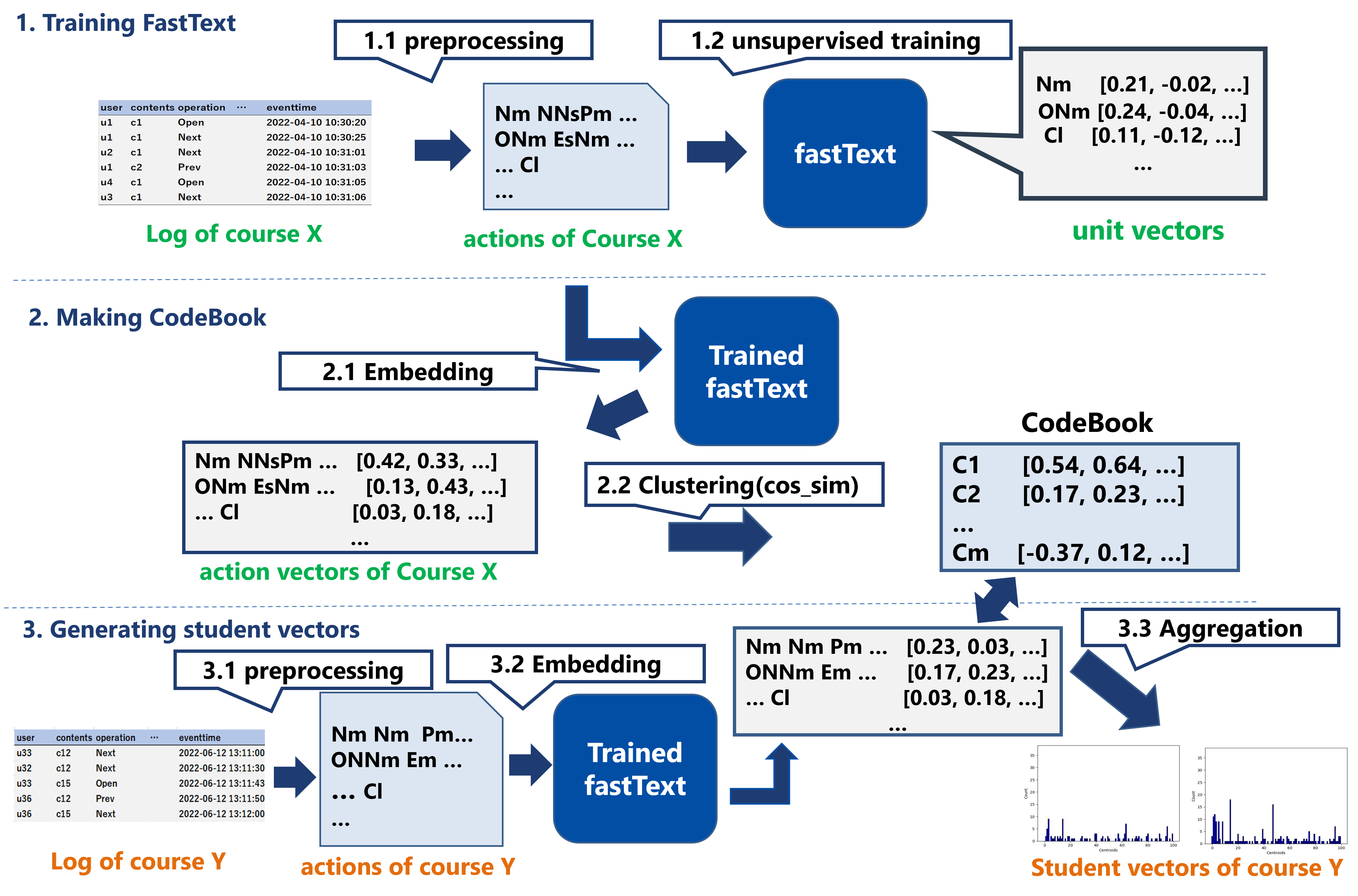}
\caption{Overview of E2Vec.}
\label{overview_of_E2Vec}
\end{figure}

This section explains how we generate distributed representations of a student's learning activity using EventStream.
\Cref{overview_of_E2Vec} illustrates our method. There are three main modules: Preprocessing, Embedding, and Aggregation. We call the embedding by these series of processes ``E2Vec,'' and use fastText \cite{bojanowski2016enriching} to acquire an embedding model. 
The details of each module are explained in the following sections. 

\subsection{BookRoll: Reading Data Collection Tool}
\label{BookRoll: Data Collection Tool}

We used an e-book system called BookRoll \cite{ogata2015book, bookroll-2018} to collect EventStream logs. Students access BookRoll to read a learning material via a web browser. There are various function buttons whose usage is recorded, such as moving to the next or previous page or adding markers.

To identify frequently occurring operations, we recorded 5,698,558 logs in BookRoll from 2021/04/01 to 2022/03/31 in various courses. \Cref{Operations of BookRoll} shows the frequently used operations; operations with less than 10,000 occurrences are grouped under ``OTHERS''. The log data indicate when, by whom, and on which material the operation was performed (see \Cref{Sample in EventStream}). The dataset used to determine these operations differs from datasets in \Cref{Section:Datasets} used to evaluate our proposed method.

\begin{table}[!ht]
\centering 
\caption{Operations in BookRoll and their correspondence to the $primitive$ symbols.}
\label{Operations of BookRoll}
\scalebox{0.85}{
\begin{tabular}{llc}
    \hline
    Operation name & Function & $Primitive$ \\
    \hline
    \textbf{N}EXT & move to the next page & $\mathrm{N}$ \\
    \textbf{P}REV & move to the previous page & $\mathrm{P}$ \\
    \textbf{O}PEN & open textbook & $\mathrm{O}$ \\
    \textbf{A}DD MARKER & draw a marker & $\mathrm{A}$ \\
    \textbf{C}LOSE & close textbook & $\mathrm{C}$ \\
    PAGE \textbf{J}UMP & go to the specified page & $\mathrm{J}$ \\
    \textbf{G}ET IT & feedback on understanding & $\mathrm{G}$ \\
    & the page contents & \\
    OTHERS & low-frequent operations & $\mathrm{E}$ \\
    \textbf{s}hort interval & 1 to 10 seconds interval & $\mathrm{s}$ \\
    & between two operations & \\
    \textbf{m}edium interval & 10 to 300 seconds interval & $\mathrm{m}$ \\
    \textbf{l}ong interval & over 300 seconds interval & $\mathrm{l}$ \\
\end{tabular}
}
\end{table}

\begin{table*}[t]
\centering
\caption{Sample of EventStream of a student.}
\label{Sample in EventStream}
\scalebox{0.85}{
\begin{tabular}{cccccccc}
    \hline
    userid & contentsid & operationname & pageno & marker & memo length & devicecode & eventtime \\
    \hline
    u1 & c1 & OPEN & 1 & & 0 & pc & 2022-04-06 13:00:00 \\
    u1 & c1 & NEXT & 1 & & 0 & pc & 2022-04-06 13:00:10 \\
    u1 & c1 & NEXT & 2 & & 0 & pc & 2022-04-06 13:00:24 \\
    u1 & c1 & NEXT & 3 & & 0 & pc & 2022-04-06 13:00:24 \\
    u1 & c1 & PREV & 3 & & 0 & pc & 2022-04-06 13:01:22 \\
    u1 & c1 & ADD MARKER & 2 & marked text & 0 & pc & 2022-04-06 13:01:30 \\
    u1 & c1 & NEXT & 2 & & 0 & pc & 2022-04-06 13:14:21 \\
\end{tabular}
}
\end{table*}


\subsection{Preprocessing}
\label{Preprocessing}

During the preprocessing phase, EventStream is converted into string representations. 
In NLP, a word is composed of several characters, a sentence consists of several words, and several sentences form documents.
To apply fastText to EventStream data, we define $primitive$, $unit$ and $action$ corresponding to ``character,'' ``word,'' and ``sentence,'' in NLP, respectively.  
An $action$ is a sequence of several $unit$s. Each $unit$ consists of a series of several $primitive$s. A $primitive$ corresponds to single character (a symbol) defined in \Cref{Operations of BookRoll}.

\Cref{Definitions} shows the definition and examples of $primitve$, $unit$ and $action$. 
Preprocessing is done by the following rules:
\begin{enumerate}
    \item EventStream log is divided by the student ID and by the content ID.     
    Therefore, a $unit$ is made from operations of a student on a lecture material. 
    
    \item Frequent operations (see \Cref{Operations of BookRoll}) are converted into the corresponding $primitive$ symbols ($\mathrm{N},\mathrm{P},\mathrm{O},\mathrm{A},\mathrm{C},\mathrm{J},\mathrm{G} \in P$); other operations are all replaced by $\mathrm{E} \in P$. 
    
    \item Insert $\mathrm{s},\mathrm{m},\mathrm{l} \in P$ between two $primitive$s, corresponding to time interval between two operations. If the interval between two operations is less than 1 second, none of the interval symbols are inserted. 
    
    \item A $unit$ ($u$) is a sequence of $primitive$s consisting of \\
    EventStream log up to 1 minute long.
     
    \item Maximal length of a $unit$ is 15 $primitive$s. If there are 15 or more $primitive$s within 1 minute, append '\_' to tail of the $unit$ and treat 15-th $primitive$ as head of the next $unit$. If 15-th $primitive$ is $\mathrm{s},\mathrm{m},\mathrm{l} \in P$, treat 16-th $primitive$ as the head of the next $unit$.
    
    \item An $action$ ($a$) is a sequence of several $unit$s.
    
    \item If time interval of two operations exceeds 5 minutes, the $action$s are separated. That is, the preceding $unit$ is the tail of the current $action$ followed by the next $action$ which begins with a new $unit$.
\end{enumerate}

\begin{table*}[t]
\centering
\caption{Definitions of elements.}
\label{Definitions}
\scalebox{0.85}{
\begin{tabular}{ccc}
    \hline
    word & definition & example \\
    \hline
    $primitive$ & an operation or a time interval between two operations & $\mathrm{N}$, $\mathrm{P}$, $\mathrm{s}$ \\
    $unit$ & sequence of $primitive$s in short span or sequence of 15 $primitive$s and '\_' & $\mathrm{Nm}$, $\mathrm{PsAl}$, $\mathrm{NNNs...N\_}$ \\
    $action$ & sequence of $unit$s & $\mathrm{Nm\  PsAl\ ...}$ \\
\end{tabular}
}
\end{table*}

To acquire the embedding of learning activities in a short time, we designed an $unit$ to contain $primitive$s for up to 1 minute or 15 $primitive$s. The maximum length setting prevents our preprocessing from generating too long $unit$s, as the percentage of $unit$s reaching max length (15) was increased when using 3 or 5 minutes instead of 1 minute. To characterize a learning activity sequence consisting of a series of $unit$s, $action$s were separated when long intervals were observed between two successive operations.

Based on the preprocessing, EventStream of a lecture course was converted into sequences of $action$s student by student. For instance, the EventStream in \Cref{Sample in EventStream} was converted into the following $action$ sequence $A_{sample}$. 
\begin{center}
$A_{sample} =\  \{\uwave{\{\uline{\mathrm{OsNmNNm}},\hspace{1mm} \uline{\mathrm{PsAl}}\}},\hspace{1mm} \uwave{\{\mathrm{N}, \ldots \}}, \ldots\} $
\end{center}
\rightline{$\uline{\ \ \ \ }: unit\hspace{3mm} \uwave{\ \ \ \ }: action$} 
The first $action$ consists of two $unit$s: $OsNmNNm$ and $PsAl$. Then, a new $action$ which begins with $primitive$ ``N'' is generated as there is an interval longer than 5 minutes between the event of ``ADD MARKER'' and subsequent ``NEXT'' (see \Cref{Sample in EventStream}).  
Note that the number of $action$s is different among students.
Some students have up to 400 $action$s representing their learning activities.
 

\subsection{Embedding}
\label{embedding}

We used fastText, a well-known model for word embedding in NLP, to generate distributed representation of $action$s. FastText generates not only the embedding vectors of words learned thus far but also unknown words. The advantage of using the fastText model is that it can generate similar vectors for two syntactically similar words. Although the fastText model is widely used and many pretrained models are publicly available, our proposed $action$ sequences have meanings different from those of natural languages. Therefore, we trained fastText with $action$s from EventStream. 
\subsubsection{Training fastText}
We followed the training strategy of fastText model based on skipgram negative sampling \cite{mikolov2013distributed}. Let $A= \{a_1, a_2, ...,  a_n$\} be a set of $action$s generated from EventStream in the training dataset. As explained earlier, an $action$ consists of a sequence of $unit$s such that each $unit$ is also an element in the set of $A$. After training the fastText model, distributed representations for each $unit$ in $A$ are obtained. In our implementation, we set the number of dimensions of the embedding vectors to 100. Regarding the other configurations, we followed the default settings in Meta Research Python module~\cite{fasttext-module} except for the minimal number of appearance parameter and epoch. We set the parameter to 1, whereas the default setting was 5, and epoch was set to 30. 

\subsubsection{Generating Action Vectors}
The trained fastText model converts a given $unit$ into its embedding representation. 
Let $\bm{u_i}$ be an embedding vector of an $unit$. The action vector $\bm{v_a}$ is calculated by using \Cref{f:action_vec}.
\begin{equation}\label{f:action_vec}
    \bm{v_{a}} =  \frac{1}{m} \sum_{i=1}^m \frac{\bm{u_i}}{|\bm{u_i}|}
\end{equation}
Note that $\bm{v_a}$ is generated student-by-student, and the number of dimensions is the same as that of $\bm{u_i}$. Therefore, we can obtain an embedded representation corresponding to each student's learning activity, $action$. 
\subsection{Aggregation}
\label{Aggregation}

Learning activities of a student over a specific period are represented by a set of $action$s. We introduce an idea of ``Bag of Words''~\cite{zhang2010understanding} to obtain a feature representation of the entire learning activity during the period, namely ``Bag of Actions.'' This step consists of creating a CodeBook using the data for training the fastText model. 
\subsubsection{Making a CodeBook}
First, it is necessary to create a CodeBook using the dataset $A = \{a_1, a_2, ..., a_n\}$ used for training the fastText model.
\begin{enumerate}
    \item Remove duplicate $action$s in $A$ and treat it as $A'$.
    \item Generate embedding vectors for all $action$s in $A'$.
    \item Perform k-means++ clustering \cite{arthur2007k} for the embedding vectors.
    \item Store the centroid of each cluster as a CodeWord, which becomes an element of CodeBook.
\end{enumerate}
The similarity between vectors was measured using cosine similarity. 

\subsubsection{Bag of Actions}

\begin{figure}[!ht]
\Description{Aggregation process of student actions.}  
\centering
\includegraphics[width=0.8\linewidth]{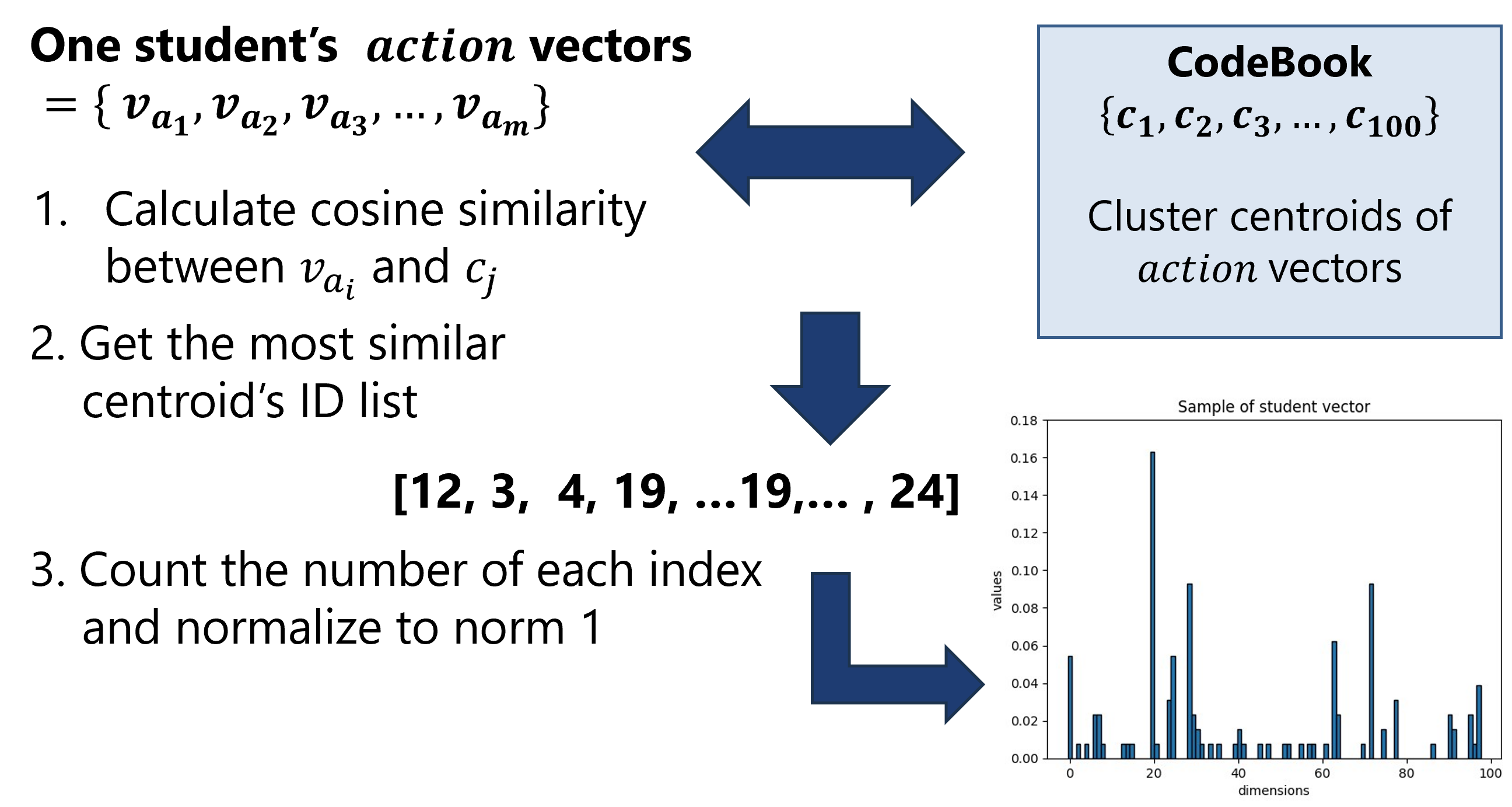}
\caption{Aggregation process of student actions.}
\label{the method of Aggregation}
\end{figure}

This is the last step needed to generate student vectors by E2Vec. Inspired by the Bag-of-Visual-Words approach \cite{Csurka2004Bagof} used in computer vision research, the proposed approach generates a histogram representation of actions based on the CodeBook. \Cref{the method of Aggregation} shows the process to aggregate one student's action vectors as corresponding student vector.

In summary, a student vector represents the characteristics of a student's learning activities as a histogram of actions over time. Its elements reflect operational patterns in the learning materials. Note that, in the proposed method, temporal information is considered in $action$ and $unit$. Therefore, we can expect to obtain a more detailed representation of learning activities than by simply aggregating the number of events, as in previous studies.

\section{Datasets}
\label{Section:Datasets}
We used EventStream logs of 6 courses that used BookRoll and students' grade information.  
\Cref{course-info} describes each course. The notations A and D represent the type of courses, and 2020, 2021, and 2022 represent the years in which the course was held. These courses were held for computer science course students in Kyushu University. 

The duration of each lecture session was 90 min. Course A is held in two consecutive lectures in one day, thus we described the lecture time of course A as 180 min. Logs recorded in and out of lecture time are used without distinction. ALL-2020 is a concatenation of A-2020 and D-2020, which were used only for training fastText and making the CodeBook, not for predicting grades of students in A-2020 and D-2020.

\begin{table}[!ht]
\centering
\caption{Information about the courses.}
\label{course-info}
\scalebox{0.9}{
\begin{tabular}{ccccc}
    \hline
    Course & Students & Logs & Weeks & Minutes \\
    \hline
    A-2020 & 60 & 142,754 & 7 & 180 \\
    A-2021 & 54 & 130,330 & 8 & 180 \\
    A-2022 & 52 & 197,389 & 8 & 180 \\
    D-2020 & 69 & 253,597 & 14 & 90 \\
    D-2021 & 106 & 287,073 & 15 & 90 \\
    D-2022 & 93 & 282,478 & 16 & 90 \\
    ALL-2020 & 129 & 396,351 & -- & -- \\
\end{tabular}
}
\end{table}

\Cref{course_grade} shows the distribution of student grades in four courses that were used as training or test data of at-risk prediction. Five grades are possible, from A to F. F means failing a course, while others pass a course. In our study, we treated grades A and B as no-risk, and C, D, F as at-risk. 

\begin{table}[!ht]
\centering
\caption{Distribution of students' grades.}
\label{course_grade}
\scalebox{0.9}{
\begin{tabular}{cccccc|cc}
    \hline
    Course & A & B & C & D & F & No-risk & At-risk \\
    \hline
    A-2021 & 9 & 11 & 10 & 18 & 6 & 20 & 34 \\
    A-2022 & 17 & 6 & 5 & 22 & 2 & 23 & 29 \\
    D-2021 & 60 & 3 & 6 & 4 & 33 & 63 & 43 \\
    D-2022 & 50 & 10 & 8 & 8 & 17 & 60 & 33 \\
\end{tabular}
}
\end{table}

We collected the data after the review of the university’s Ethical Committee. In addition, students were offered that their data can be used for research, and they declared their intention to opt-in or opt-out for this use of data.

\section{UNIT vector analysis (RQ1a)} 
\label{P: Analysis about UNIT vectors}
In this study, we verify which $unit$s are similar to each other in the vector representation and find similarities and differences between the vectors of known and unknown $unit$s. To generate $unit$ vectors, we used fastText trained with $A_{ALL-2020}$. fastText learned 33963 $unit$s in $A_{ALL-2020}$.  

\subsection{Method}
\label{M: Analysis about action vectors}
As an example demonstration, we calculated similarity between $Nm$ -- the most frequent unit generated from students' activity -- and all the $unit$s in $A_{ALL-2020}$. For comparison, the same calculation was performed for $NNNNsNmNsNsPl$ -- unlike $Nm$, this $unit$ was not trained by fastText. The former $unit$ is in $A_{ALL-2020}$ and $A_{D-2022}$. The latter is not in $A_{ALL-2020}$, but is included in $A_{D-2022}$.

In addition, we calculated cosine similarity between each of two $unit$s and all the $unit$s in $A_{D-2022}$ and evaluated the distribution of cosine similarities.

\subsection{Result}
\label{R: Analysis about action vectors}

\Cref{Similar actions of "Nm"} and \Cref{Similar actions of $NNNNsNmNsNsPl$} shows the $unit$ similarity results. E2Vec generated embedding vectors with high similarity ($\geq$ 0.7) when $unit$s had common subwords -- i.e., high-similarity $unit$s have common subwords (subword is a sequence of $primitive$s). $Nm$ and its highly similar $action$s share the subword $Nm$.
A common subword implies that two $unit$s are created for the same order of operations and similar time intervals.  

\begin{figure}[!ht]
\Description{Histogram of cosine similarity between $Nm$ and 25078 units in D-2022}
\centering
\includegraphics[width=0.7\linewidth]{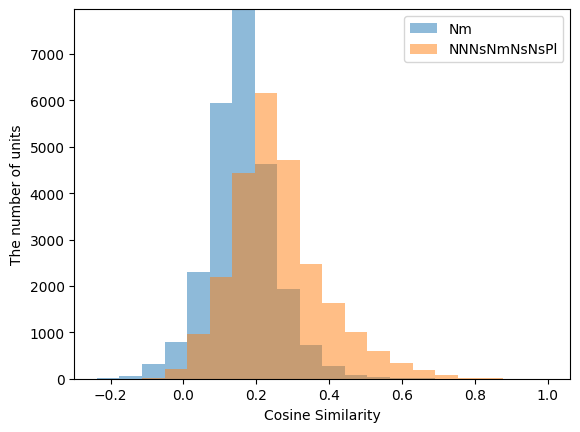}
\caption{Histogram of cosine similarity between $\mathrm{Nm}$ or $\mathrm{NNNNsNmNsNsPl}$ and 25076 units in D-2022.}
\label{histogram of cosine similarity Nm NNNNsNmNsNsPl}
\end{figure}

In addition, \Cref{histogram of cosine similarity Nm NNNNsNmNsNsPl} is a histogram of the similarity between the two selected units and all 25076 $unit$s in D-2022. One $unit$ has fewer similar units and many dissimilar units in the course activities of students on the learning materials. This shows that not only features with high similarity but also features with low similarity were generated.

\begin{table}[!ht]
\centering
\caption{The most similar units to $\mathrm{Nm}$.}
\label{Similar actions of "Nm"}
\begin{tabular}{cc}
    \hline
    $unit$ & similarity \\
    \hline
    $\mathrm{NmONm}$ & 0.835 \\
    $\mathrm{NmOsCNm}$ & 0.734 \\
    $\mathrm{NmGPNm}$ & 0.706 \\
    $\mathrm{NmENm}$& 0.703 
\end{tabular}
\end{table}

\begin{table}[!ht]
\centering
\caption{The most similar units to $\mathrm{NNNNsNmNsNsPl}$.}
\label{Similar actions of $NNNNsNmNsNsPl$}
\begin{tabular}{cc}
    \hline
    $unit$ & similarity \\
    \hline
    $\mathrm{NNNNsNmNsNsPm}$ & 0.937 \\
    $\mathrm{NNNsNmNsPl}$ & 0.894 \\
    $\mathrm{NNNsNmNs}$ & 0.863 \\
    $\mathrm{NNNNNsNmNs}$ & 0.855 
\end{tabular}
\end{table}

Overall, these results suggest that the similarity between learning activities in a short time is preserved in distributed representations of $unit$s generated with fastText. Also, dissimilar activities converted into discriminative features. This is a property of a good feature generator, which should produce highly similar features for similar inputs and highly discriminative features for those that are not.

\section{ACTION vector analysis (RQ1b)}
\label{P: Analysis about centroids in CodeBook}
A vector of an $action$ $a = u_1,\ u_2,\ ... \ , u_m$ is generated from $unit$ vectors $\{\bm{u_1}, \bm{u_2}, ..., \bm{u_m}\}$. 
In our method, the CodeBook is made to aggregate one student's $action$s to student vector. In the process of making the CodeBook, k-means++ clustering is performed with the set of $action$ vectors. In this section, we describe the study of $action$ vectors. We evaluated the result of clustering and sought to find the features of $action$ vectors. 

\subsection{Method}
\label{M: Analysis about centroids in CodeBook}
We used fastText, which was trained using $A_{ALL-2020}$. It has 19979 $action$s, of which 14016 were unique.
In this study, k-means++ clustering was performed using $A_{ALL-2020}$. The number of clusters $k$ was set to 10. 

We calculated the maximum, mean, and variance of the length of $action$s in each cluster.
The length of $action\ a$ is, in other words, the number of $unit$s that constitute $a$. A long sequence of $unit$s indicates that students execute sequential operations on learning materials in BookRoll.  

\subsection{Result}
\label{R: Analysis about centroids in CodeBook}

\Cref{Statics of clusters} shows the quantitative analysis of clusters. Cluster numbers are sorted by max length of $action$s. From the $action$s length in each cluster, $c_0$ and $c1$ contains $action$s that consist of small number of $unit$s. 
The maximum length $action$ in this cluster had only 8 and 9 $unit$s. Other clusters involved $action$s, which consists of large number of $unit$s.
However, the mean of the length of $action$s in each cluster was less than 10. 
Also, It can be observed that the $action$s in the same cluster have similar $unit$ and sequences of $unit$s. For example, one cluster have sequences composed only of multiple $N$s. These sequences indicated that the student clicked the NEXT multiple times per second. On the other hand, some $action$s do not have similar $unit$s to other $action$s in the same cluster; and similar $action$s in different clusters.

\begin{table}[!ht]
\centering
\caption{Descriptive statistics of clusters.}
\label{Statics of clusters}
\scalebox{0.71}{
\begin{tabular}{ccccccccccc}
    \hline
    cluster & $c_0$ & $c_1$ & $c_2$ & $c_3$ & $c_4$ & $c_5$ & $c_6$ & $c_7$ & $c_8$ & $c_9$ \\
    \hline
    max & 8 & 9 & 19 & 23 & 24 & 30 & 33 & 34 & 37 & 64 \\
    mean & 1.8 & 2.0 & 4.0 & 3.3 & 3.2 & 3.5 & 3.9 & 7.0 & 5.0 & 6.3 \\
    variance & 1.04 & 0.93 & 9.91 & 6.78 & 5.67 & 8.04 & 9.40 & 28.6 & 24.9 & 37.7 \\
    \#$action$ & 1261 & 1471 & 818 & 1280 & 1139 & 1063 & 915 & 1859 & 1118 & 3092 \\
\end{tabular}
}
\end{table}

\section{At-risk prediction (RQ2, RQ3)}
\label{P: At-risk Prediction}
In this study, we performed at-risk prediction as an application of student vectors generated by E2Vec. This is a binary classification task. In our setting, classification models trained with all the student vectors and their grades of $c_x$, predicted all the students' grades of course $c_y$ from the student vectors of $c_y$. The ground truth of this task is student's final grade of $c_y$. At-risk are treated as positive and no-risk as negative classes. 
This evaluation aims to verify the following hypotheses. 
\begin{itemize}
    \item[h1:] The student vectors generated by E2Vec are effective as an input of classification models for at-risk prediction.
    \item[h2:] fastText trained with logs from several courses is able to perform predictions for various courses.  
    \item[h3:] Accurate predictions can be made regardless of the combination of train and test lecture courses by using the proposed method.
\end{itemize}
h1 corresponds to RQ2. h2 and h3 are related to RQ3. 

\subsection{Method}
\label{M: At-risk Prediction}

\begin{figure}[!ht]
\Description{At-risk prediction process.}  
\centering

\includegraphics[width=0.82\linewidth]{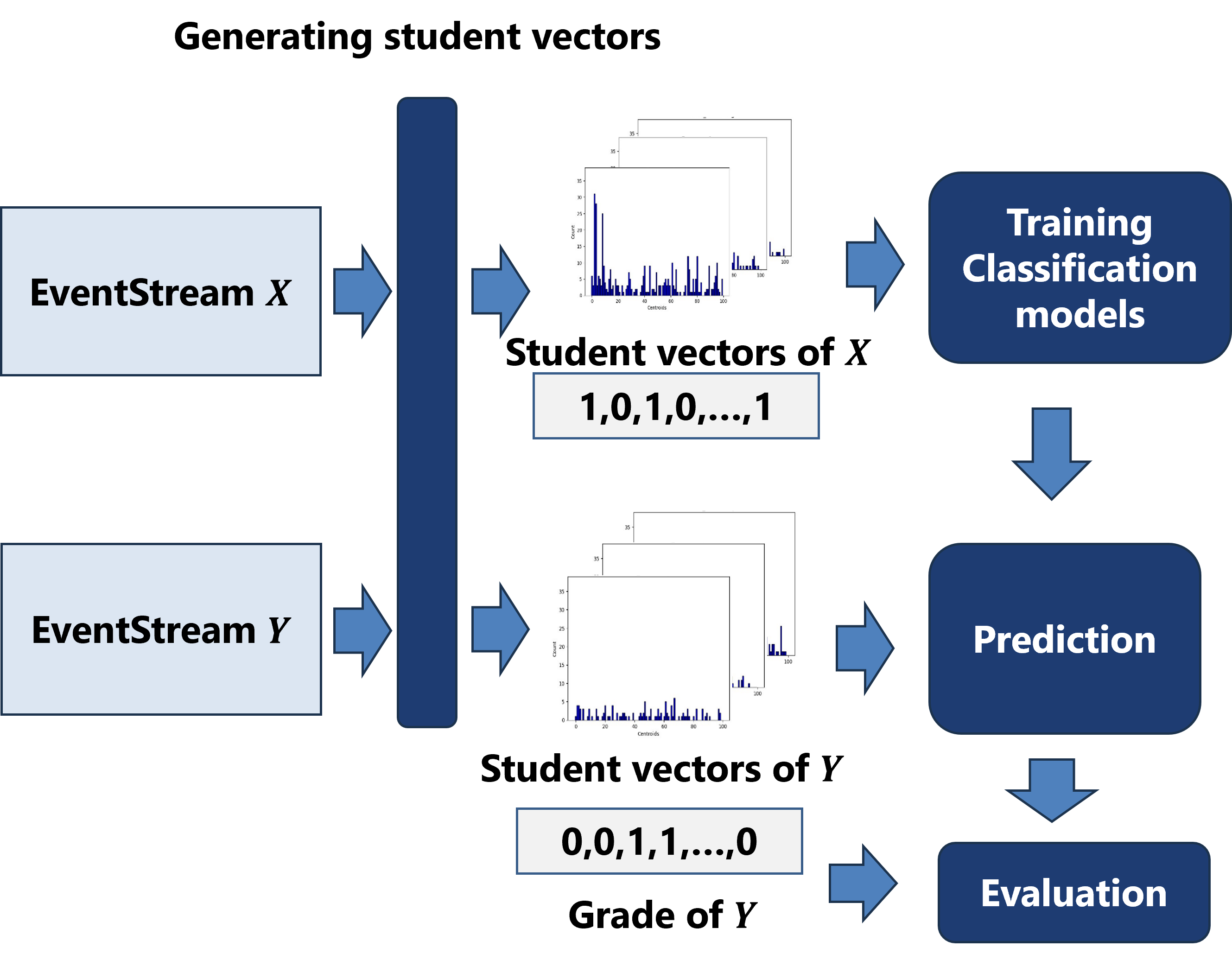}
\caption{At-risk prediction process (the graphs only illustrate different distributions; the values are not expected to be read).}
\label{the process of At-risk prediction}
\end{figure}

We used four classification models: Random Forest Classifier (RFC), Support Vector Machine (SVM), Ada-Boost classifier (ADA), and k-Nearest Neighbor classifier (KNN), all from the Scikit-learn library \cite{pedregosa2018scikitlearn}. We aimed to verify that E2Vec features are suitable as input of different classification models. Two courses are selected from the four courses in \Cref{course_grade}. We represent the training course as $X$ and test course as $Y$. The at-risk prediction process involves three steps: the generation of student vectors, training of the classification model, and prediction. This process is illustrated in \Cref{the process of At-risk prediction}.

\subsubsection{The Generation of Student Vectors}
\label{Generating student vectors}
We used the proposed method and two comparison methods to generate vectors of all students in $X$ and $Y$, and used these vectors as input of classification models.

\begin{enumerate}
    \item E2Vec\\
    fastText trained with ALL-2020 was treated as basic model and compared models trained with A-2020 and D-2020 used to verify h2.  
    The number of centroids $k$ was selected from $[10,100]$. $k$ was corresponds to the size of CodeBook and student vector. The compared models' dimensions were set to 100.
    
    
    \item Operation Count Feature (OC)\\
    One student vector was formed by counting the number of each operation in \Cref{Operations of BookRoll} and normalizing to norm~1. Thus, the dimension of student vectors were 7.
\end{enumerate}

\subsubsection{Training Classification Model}
\label{Training Classification model}
To tune the hyperparameters of the classification model, we performed a grid search with 3-fold cross validation with training data. This was implemented using the scikit-learn module \texttt{GridSearchCV}. Students in $X$ were split into three sets, two of which were treated as training data, and the other as validation data. The score of the one-hyperparameter set model was the mean of 3 F1-scores in each validation. The parameter set with the highest score was selected. In the prediction, the selected model and the default parameter model were used. In our implementation, the random states of the models were fixed at 42 to ensure reproducibility. 

\subsubsection{Prediction}
\label{Prediction}
In this step, two trained classification models (the selected parameter set model and the default parameter model) predicted all students in $Y$ as at-risk or not with the student vector. The output of the classification model is a binary list of labels. Label 1 indicates that the student is predicted to be at-risk. For the evaluation, we calculated F1-score using the predicted labels and ground truth. \Cref{T:Result of At-risk Prediction_A} to \ref{T:Result of At-risk Prediction_RFC} show only the higher F1-score of each model's (tuned hyperparameters model and default hyperparameters model) evaluation.

\subsection{Result}
\label{R: At-risk Prediction}

\Cref{T:Result of At-risk Prediction_A} shows the result of prediction for A-2022, trained on A-2021. \Cref{T:Result of At-risk Prediction_D} represents prediction results of D-2022, with classifiers trained on D-2021. These are the cases when the model was trained with past student information and predicted students on the same course in the following years. This is a typical application of at-risk prediction.

\begin{table}[!ht]
\centering
\caption{Prediction F1-scores. Train: A-2021, Test: A-2022.}
\label{T:Result of At-risk Prediction_A}
\scalebox{0.73}{
\begin{tabular}{c|ccccc}
    \hline
    models & E2Vec$_{k10}$ & E2Vec$_{k100}$ & E2Vec$_A$ & E2Vec$_D$& OC\\
    \hline
    RFC & 0.68 & \textbf{0.72} & 0.68 & 0.71 & 0.60\\
    SVC & 0.70 & 0.69 & \textbf{0.72} & 0.61 & \textbf{0.72} \\
    ADA & \textbf{0.63} & 0.60 & 0.62 & 0.54 & 0.57 \\
    KNN & \textbf{0.63} & 0.49 & 0.54 & 0.42 & \textbf{0.63}\\
\end{tabular}
}
\end{table}

\begin{table}[!ht]
\centering
\caption{Prediction F1-scores. Train: D-2021, Test: D-2022.}
\label{T:Result of At-risk Prediction_D}
\scalebox{0.73}{
\begin{tabular}{c|ccccc}
    \hline
    models & E2Vec$_{k10}$ & E2Vec$_{k100}$ & E2Vec$_A$ & E2Vec$_D$& OC \\
    \hline
    RFC & \textbf{0.65} & 0.64 & 0.63 & 0.61 & 0.63 \\
    SVC & 0.59 & \textbf{0.62} & \textbf{0.62} & 0.60 & 0.61 \\
    ADA & \textbf{0.68} & \textbf{0.68} & 0.63 & 0.62 & 0.56 \\
    KNN & 0.45 & 0.35 & 0.30 & 0.35 & \textbf{0.56} \\
\end{tabular}
}
\end{table}

In these typical cases, the proposed E2Vec method yielded equivalent or higher F1-scores compared to Operation Count Feature, which scored better only in one case (KNN for course D). Additionally, E2Vec$_{k10}$ and E2Vec$_{k100}$ were trained using ALL-2020. The prediction results using these models were not much lower than those of E2Vec$_A$ and E2Vec$_D$ trained with only A-2020 and D-2020. The predicted A-2022 students with RFC and E2Vec$_{k100}$ scored 0.72. It achieved a better F1-score than E2Vec$_A$ (0.68).  

Therefore, for verifying h1, the proposed method is effective when used as the embedding method for at-risk prediction. Also, for h2, fastText trained with several courses can be used without declining of prediction quality compared to when trained with one course. 

\Cref{T:Result of At-risk Prediction_RFC} shows the result of prediction with RFC for all combinations of training and test courses. The sets of training and test data are represented as (Train, Test). As a result, in most cases, predictions using the same course data such as (A-2021, A-2022), (A-2022, A-2021), (D-2021, D-2022) and (D-2022, D-2021) achieved high F1-scores. In particular, the F1-score of the prediction for (D-2022, D-2021) with E2Vec$_{k100}$ is 0.85. 
However, the predictions for some datasets had lower F1-scores. For example, prediction of (D-2021, A-2022) with E2Vec$_{k100}$ scored 0.24. It was much lower than the Operation Count Feature's score of 0.73.

\begin{table}[!ht]
\centering
\caption{F1-score of at-risk Prediction with RFC.}
\label{T:Result of At-risk Prediction_RFC}
\scalebox{0.73}{
\begin{tabular}{cc|ccccc}
    \hline
    Train & Test & E2Vec$_{k10}$ & E2Vec$_{k100}$ & E2Vec$_A$ & E2Vec$_D$& OC \\ \hline
    A-2021 & A-2022 & 0.68 & \textbf{0.72} & 0.68 & 0.71 & 0.60 \\
           & D-2021 & 0.55 & \textbf{0.60} & 0.58 & 0.59 & 0.52 \\
           & D-2022 & 0.49 & \textbf{0.53} & \textbf{0.53} & \textbf{0.53} & 0.50 \\ \hline
    A-2022 & A-2021 & 0.69 & \textbf{0.71} & 0.69 & 0.70 & 0.63 \\
           & D-2021 & 0.69 & 0.67 & \textbf{0.76} & 0.75 & 0.68 \\ 
           & D-2022 & 0.48 & 0.51 & 0.59 & \textbf{0.61} & 0.36 \\ \hline
    D-2021 & A-2021 & 0.63 & 0.53 & 0.41 & 0.41 & \textbf{0.74} \\
           & A-2022 & 0.63 & 0.24 & 0.24 & 0.24 & \textbf{0.73} \\
           & D-2022 & \textbf{0.65} & 0.64 & 0.63 & 0.61 & 0.63 \\ \hline
    D-2022 & A-2021 & 0.54 & 0.59 & 0.56 & 0.56 & \textbf{0.74} \\
           & A-2022 & 0.47 & 0.38 & 0.41 & 0.29 & \textbf{0.67} \\
           & D-2021 & 0.80 & 0.85 & 0.85 & \textbf{0.86} & 0.81 \\
\end{tabular}
}
\end{table}
Based on these results, training data of the classifiers affected the result of prediction using feature vectors generated by the proposed model. In other words, h3 was rejected. When training data and test data are the same course held in different year such as (D-2021, D-2022), F1-score of at-risk prediction is higher than in most other cases. Whereas, when using data of course D- for training and A-2021 or A-2022 for testing, E2Vec achieved much lower F1-scores than OC.

\section{Discussion}
\label{Discussions}

For RQ1, the effectiveness of E2Vec was shown in \Cref{P: Analysis about UNIT vectors} and \Cref{P: Analysis about centroids in CodeBook}. The $unit$ vectors preserve the similarity of student operations in a short time. Regardless of whether $unit$ is involved in the vocabulary of fastText, the input $unit$ and the similar $unit$s shared sequences of $primitive$s. In addition, $action$ vectors have similarities that correspond to the sequences of operations. Based on the clustering, $action$s in the same cluster have similar $unit$s. Thus, the distributed representations of $unit$s generated by E2Vec are effective representations of student learning activities. 

For RQ2, \Cref{P: At-risk Prediction} showed that E2Vec can be used as a feature expression method for at-risk prediction using several ML models. We investigated a practical use case of using past course data for model training and the following year's data for prediction. In typical cases, the F1-scores using the features generated by E2Vec were comparable or higher than the traditional approach, so we conclude that the features can be suitably utilized for downstream tasks.

For RQ3, F1-scores of using E2Vec$_{k100}$, in which mixed course data was used to train fastText, were comparable to those of using single course data such as E2Vec$_A$ or E2Vec$_D$ (see \Cref{T:Result of At-risk Prediction_A} and \Cref{T:Result of At-risk Prediction_D}). In general, using the same course data used in feature extraction and its prediction task led to successful results. On the other hand, our results suggest that fastText model does not have to be trained by specific course data. Once a fastText model is trained using a mixed dataset from past courses, the model can extract robust features that can be used universally. When we used E2Vec features for at-risk prediction, the training data used for the classification model (not fastText model) exhibited differences of prediction quality. It implied that the same learning activity has a different influence on the final grade in each course. 

The limitation is that our study used data from six courses at our university. The number of students in one course ranged from 50 to 100, and the grade distributions differed. In addition, if our method is used with other e-book EventStream, additional preprocessing is required, because our implementation only corresponds to BookRoll EventStream.

\section{Conclusion}
\label{concludions}
We proposed E2Vec -- a method of feature expression from e-book EventStream. Our method uses fastText, a word embedding model, to learn the distributed representations of operations in a short time, called $unit$. Similar $unit$s have similar representations, and $action$s in the same cluster have a common $unit$s or subword. We applied our method to at-risk prediction, which is a representative task in EDM. Our model recorded a higher F1-score than the operation count features for at-risk prediction in typical cases.

Future work should apply E2Vec with deep learning models or other EDM tasks, such as early prediction of student dropout. Finally, the impact of changing intervals on the prediction performance needs to be evaluated.

\section{Acknowledgments}
This work was supported by JST CREST Grant Number JPMJCR22D1 and JSPS KAKENHI Grant Number JP22H00551, Japan.

\bibliographystyle{abbrv}
\bibliography{references}  

\balancecolumns

\appendix

\section{Supplementary Materials}
\label{appendix:materials}

The code written to produce the results reported in this paper is publicly available at:\\
\url{https://github.com/limu-research/2024-edm-e2vec}.

\section{OpenLA}
\label{appendix:openla}

E2Vec functions have been implemented in ``OpenLA: Library for Efficient E book Log Analysis and Accelerating Learning Analytics''~\cite{murata2020openla}, which is available at 
 \url{https://limu.ait.kyushu-u.ac.jp/~openLA/}. The trained model used in this paper is also available at \url{https://limu.ait.kyushu-u.ac.jp/~openLA/models/}. 

\end{document}